# EXPERIENCE WITH MACHINE PROTECTION SYSTEMS AT PIP2IT *

A. Warner†, M. Austin, L. Carmichael, J.-P. Carneiro, B. Hanna, E. Harms, R. Neswold, L. Prost, R. Rivera, A. Shemyakin, M. Ibrahim, J. Wu, Fermilab, Batavia, IL 60510, USA

*Abstract*

The PIP-II Injector Test [1] (PIP2IT) facility accelerator was assembled in multiple stages in 2014 – 2021 to test concepts and components of the future PIP-II linac that is being constructed at Fermilab. In its final configuration, PIP2IT accelerated a 0.55 ms x 20 Hz x 2 mA H⁻ beam to 16 MeV. To protect elements of the beam line, a Machine Protection System (MPS) was implemented and commissioned. The beam was interrupted faster than 10µs when excessive beam loss was detected. The paper describes the MPS architecture, methods of the loss detection, procedure of the beam interruption, and operational experience at PIP2IT.

## INTRODUCTION

PIP2IT is a prototype accelerator assembled as a testbed for developing and testing some of the novel and challenging technologies required to construct the Proton Improvement Plan-II (PIP-II) project at Fermilab [2]. The central element of PIP-II is the 2mA, 800 MeV H⁻ linac which comprises a room temperature front end followed by an SRF accelerator. The PIP2IT (Fig. 1) was constructed in two phases. The first phase consisted of the room temperature portion of the machine, or Warm Front-end (WFE), which comprises a H⁻ ion source, a radio-frequency quadrupole (RFQ) and a transport line for delivering beam to the superconducting section of the accelerator at 2.1 MeV. In the second phase, it was appended by two cryomodules called a Half Wave Resonator (HWR) and a Single Spoke Resonator type1 (pSSR1).

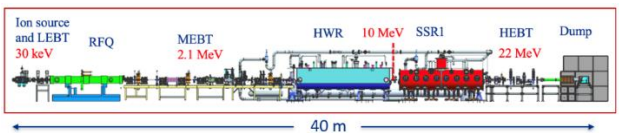

Figure 1: PIP2IT Accelerator Layout

In the final PIP2IT operational scenario, a 5 mA, 0.55 ms x 20 Hz beam out of the RFQ was scraped transversely in the Medium Energy Beam Transport (MEBT), primarily with scrapers, and then half of the bunches was removed by a bunch-by-bunch chopping system in an aperiodic pattern required for the future injection into the Booster [3]. Then, the beam with 2 mA pulse current was accelerated in the HWR and pSSR1 cryomodules to 8 and 16 MeV, correspondingly, and transported through the High Energy Beam Transport (HEBT) to a beam dump. A machine protection system capable of inhibiting the beam within 10 µs in response to an acute beam loss, protecting machine components from beam damage, and monitoring the machine configuration was developed.

## MPS SCHEME

The PIP2IT MPS scheme tested several features that had been envisioned to be used at PIP-II.
- The main layer of protection is beam inhibiting by a small number (four) of devices that were carefully tested and administratively controlled. All of them are upstream of the RFQ.
- The global protection of the linac is performed by comparison of signals from pairs of current-sensitive devices. These devices (4 pairs) are administratively controlled.
- The second layer of protection, the local protection, controlled the beam loss to multiple electrically isolated electrodes in the MEBT.
- The third level was the readiness signals from the subsystems (vacuum, RF etc.).
- Finally, the beam could be operated only at specific combinations of beam intensities and machine configurations defining how far the beam can propagate. Each combination had its own table of the MPS parameters to control.

This hierarchy, on one hand, provided a robust protection, and, on the other hand, a significant flexibility in operation.

## PROTECTION SYSTEM DESIGN

The machine protection system at PIP2IT was not a distributed system and was intended as a testbed for demonstrating the methods needed to protect accelerator components, adequately monitor beam current losses, develop the system logic, provide flexibility to beam measurements and to test hardware and tools required to build a larger scale distributed system for the PIP-II project.

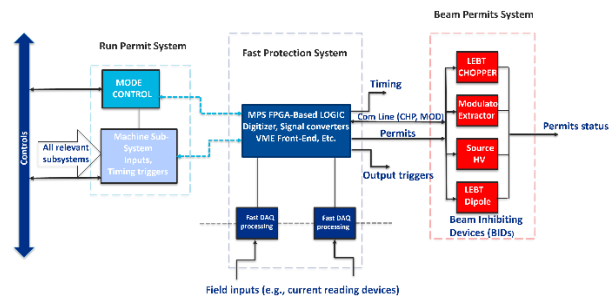

Figure 2: Simplified MPS concept.

The MPS ultimately received Ok/Not-Ok signals from subsystems and managed permits to Beam Inhibiting Devices (BIDs). Figure 2. shows the simplified block diagram. The Low Energy Beam Transport (LEBT) chopper was the primary BID that was cutting the beam off 150 ns after it's permit went away. The other three BIDs were the ion source extractor voltage and its bias voltage as well as the

LEBT bend current. While the extractor reaction time was similar to the chopper's, the bias voltage and the bend current were dropping in sub-second range.

The MPS input signals were processed at the same rate independently on the signals origin, removing the permit from BIDs within 1 μs after receiving the interruption. While an automated beam recovery after intermittent trips has been foreseen, it was not implemented because of the planned continuously changing operational conditions of the test machine.

Reaction of the MPS to the input signals was defined by the tables generated for specific accelerator conditions, described by machine configurations and beam modes.

Machine configurations determined how far beam could propagate along the beamline. The MPS was checking that the devices limiting the beam propagation were activated (e.g., the scrapers were moved in, and the valve downstream closed) and masked the irrelevant downstream sensors. Four implemented machine configurations are shown in Table 1. Implementation of the machine configurations and software to switch between them provided important flexibility in commissioning. At the start, sections of the accelerator could be sequentially commissioned without modification of the MPS. Later, two different commissioning activities could be performed in parallel. For example, with the machine in the MEBT configuration, testing of the MEBT absorber was performed in parallel with RF commissioning of the SRF cavities.

Table 1: PIP2IT MPS Configurations

| Configuration | Purpose | Beam to | Restriction |
|---|---|---|---|
| Ion Source (IS) | Condition | No beam | IS valve |
| LEBT | IS & LEBT tuning | End of LEBT | LEBT scraper |
| MEBT | MEBT Tuning | MEBT absorber | MEBT scrapers |
| Full Line | HEBT | Dump | All open |

The allowed beam power, controlled by the pulse duration, was defined by two beam modes: a diagnostic mode, with maximum pulse duration of 10 μs for tuning, commissioning, and beam studies, and an Operational mode with a maximum pulse duration up to 0.55ms consistent with PIP-II beam requirements. In the diagnostic mode, the maximum instantaneous power density was estimated to be below the damage threshold for insertable devices such as scrapers, wires, Allison scanners, and Faraday Cup. In the operational mode with much higher damage potential, the insertable devices were not allowed to be moved in. A so-called Mode Controller program managed the configurations and modes available to the user. It provided a comprehensive overview of the machine status, enforced, and monitored the machine configurations, and set limits on beam parameters (beam modes) for the defined machine configurations.

## BEAM LOSS MEASUREMENTS

The complete assortment of beam current sensing devices is shown in figure 3. Because the beam damage potential in the ion source, LEBT, and the RFQ was considered to be low, current readings from these areas are not fed into the MPS. Because of the tight space limitations in the MEBT, the devices there are compact electrostatic pickups (RPU, described later). Since part of the beam is removed in the middle of the MEBT by the chopping system, there are two RPU pairs, upstream and downstream of the MEBT absorber. The beam loss in the cryomodules is controlled by comparison of currents read by ACCTs placed at the exit of the MEBT and entrance of the HEBT, and the HEBT loss is calculated by comparison of currents of the ACCT at its entrance and the beam dump.

Measurement of the uncontrolled beam loss in the MEBT is more complicated. First, half of the bunches is removed by the MEBT chopping system onto the MEBT absorber. Therefore, the MPS controls separately the beam loss upstream and downstream of the absorber. Second, the MEBT elements are tightly packed, and inserting two additional ACCTs on both sides of the absorber is difficult. Instead, the control of the loss is made with four electrostatic pickups, so-called Ring Pickups (RPUs). Each RPU is a metal cylinder mounted inside a flange. The RPU electronics analyses the 162.5 MHz component of the capacitive signal from the cylinder and delivers its amplitude to the MPS. Importantly, this signal doesn't depend on the pulse width in the wide range from microseconds to CW. In combination with constant beam properties through the pulse, it allowed beam tuning in the MEBT with 10 μs pulses to calibrate the RPUs. With chopping off, the beam was propagated through the MEBT, and the loss was identified by comparison of ACCTs at the entrance and exit of the MEBT. In this state, the RPU signals were calibrated to take into account the effect of the beam width on the RPU signal. After tuning to scrape the beam as required and to minimize the undesired loss to acceptable levels, comparison of the signals from two RPU pairs (beginning of the MEBT to entrance of the absorber box and downstream of the absorber box to the MEBT exit) was activated in the MPS.

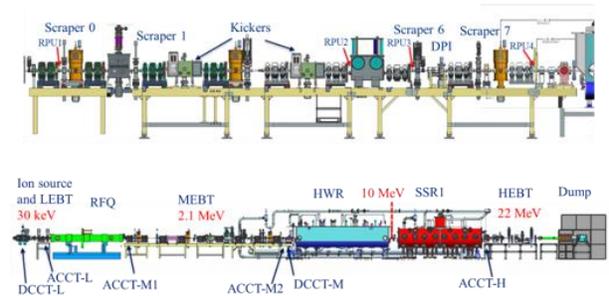

Figure 3: Layout of the Beam-Current channels

The RPUs were also used by the MPS to monitor the beam pulse length as an additional protection layer in a case of the LEBT chopper failure.

Comparison of currents with ACCTs and RPU provided the global protection of the corresponding sections. To add a protection layer for the elements restricting the beam aperture and, therefore, the most susceptible to the beam damage, these elements were electrically isolated, biased to suppress the secondary electron emission, and their signal, fed into the MPS, and were monitored for extensive beam loss. These channels included:

- 16 channels of horizontal and vertical scrapers, housed in 4 crosses. They were used to scrape the beam and configure the beamline in the MEBT mode.
- 4 Kicker Protection Electrodes (PEs), metal plates placed at both ends of the chopper deflecting structures to protect these structures from primary beam exposure.
- Differential Pumping Insert (DPI), a 200 mm long, 10 mm ID pipe which separated high vacuum and ultra-high vacuum, particle free regions.

The thresholds for current losses at these devices were included into the MPS configuration files.

*Beam loss protection integral*

The primary MPS tool used to protect the MEBT and the cryomodules at PIP2IT was based on the comparison of readings from the current measuring devices. In the initial implementation, the beam current was read in a 0.25 ms window, and the average values were compared. The bias of each individual device was adjusted based on reading the device during the beam-off time between pulses. In the final implementation, it was found to be advantageous to use an integration method with a limit on the integral moderated by a damping function. The integration improved the stability of the loss detection by decreasing the sensitivity to the measurement noise. In addition, comparison of the threshold could be made continuously instead of waiting for the end of an acquisition window. This method also improved the processing and reaction time to that of the FPGA clock speed and digitizer resolution. This algorithm was applied both to comparison of currents in the global protection and individual current signal in the local protection.

The FPGA based implementation of the algorithm instructed the MPS to drop the beam permit signal if the value $Q(T)$ calculated according to Eq. (1) exceeded a user-specified threshold $Q_{th}$:

$$Q_{th} < Q(T) = e^{\frac{-T}{\tau}} \int_0^T \Delta I(t) \cdot e^{\frac{t}{\tau}} dt \qquad (1)$$

where $T$ is the time of evaluation. Eq. (1) assumes that the integration starts before actual start of the beam and after the beam-off window for baseline subtraction. $\Delta I(t)$ is the difference between readings of two beam current devices at time $t$; $\tau$ is a time constant.

For a constant beam loss in a short time scale, $T \ll \tau$, $Q(T) \approx \Delta I \cdot T$. Correspondingly, the threshold $Q_{th}$ should be chosen to ensure that in the case of total beam loss with current $I_0$, the interruption condition is fulfilled in the time $T_{th}$ safe for the protected section:

$$Q_{th} = I_0 \cdot T_{th} \qquad (2)$$

For a constant beam loss in long time scales, $T \gg \tau$, $Q(T) \approx \Delta I \cdot \tau$. The choice of the constant $\tau$ is determined by the maximum beam loss sustainable in steady state $\Delta I_{max}$:

$$\tau = {Q_{th}}/{\Delta I_{max}} \qquad (3)$$

Such an algorithm roughly mimic's how the temperature rises in a metal irradiated by a beam, where transition from μs to ms scales corresponds to thermal conductivity playing a significant role. Numerical estimations of various loss scenarios showed that with chosen thresholds the temperature rise in the beam irradiated stays well below 1000 K. Figure 4 illustrates the concept as applied to the cryomodules. The resolution for losses was ~3% limited by noise.

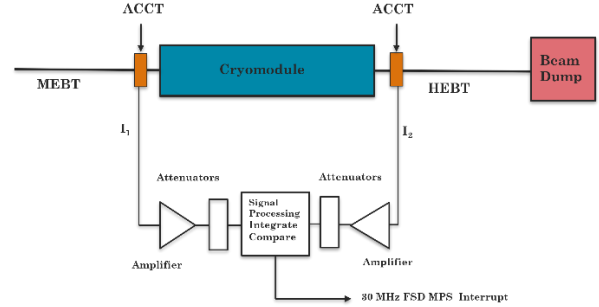

Figure 4: Cryomodule Protection Concept

Three ACCTs, four RPUs and the beam dump were incorporated into the so-called Differential Beam Current Monitoring (DBCM) system. The discrete form of the integral described was implemented on FPGA based, 125 MHz digitizers on a VME platform. Figure 5. shows the block design for the analog current signals and the VHDL implementation.

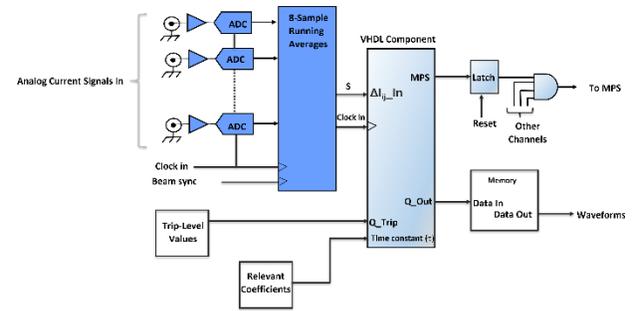

Figure 5: System Digitizer and VHDL Component

The VHDL component was designed to remove a 30 MHz MPS permit signal which in turn removes a 5 MHz chopper permit if the integration component was interrupted from calculating. The code that did the calculation made the comparisons with the limit values established in Table 2. These parameters for the various threshold limits were protected and controlled as part of a larger set of configuration parameters for the MPS. This implementation removes the MPS permit out to the main MPS logic system in 8ns if a threshold is exceeded as measure on the system clock, Figure 6.

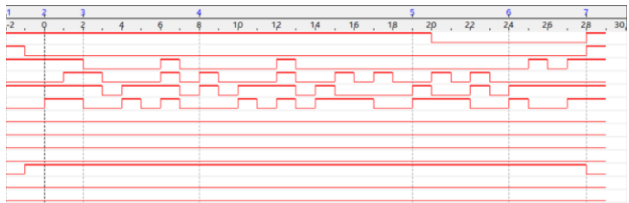

Figure 6: DBCM response time to threshold trips

Table 2: PIP2IT MPS Protection Parameters

| System | Power, W | $Q_{th}$, mA·µs | τ, µs | Based on |
|---|---|---|---|---|
| Scrapers | 75 | 100 | 5000 | Scraper current |
| HWR/p SSR1 | 20 | 20 | 2000 | ACCT diffs |
| PEs | 40 | 100 | 5000 | PE current |
| DPI | 25 | 2 | 2000 | DPI current |

## DISTRIBUTED SYSTEM HARDWARE

Part of the MPS development at PIP2IT was focused on designing and testing the hardware necessary to build a distributed MPS for protecting the future PIP-II accelerator and demonstrate methods for integration with the larger accelerator complex at Fermilab. The hardware required to concentrate subsystem signals, manage permits, and beam aborts was tested to verify signal integrities, distribution with optical fibre links, reaction timing etc. These concentrators where specifically used to manage and consolidate the signals derived from the HWR and SSR1 cryomodule Quench Protection Circuits and the RF Protection circuits. Combinations of these units were installed with appropriate cable lengths to mimic a distributed system. The units dropped the local permit within 100 ns of an input going low and sent signals to subsequent concentrator units within 140 ns to 270 ns. The signal is analyzed for integrity and passed along again repeating the sequence. The hardware Figure 7. operates in a failsafe manner with active-high TTL logic. Each units have 8 optically isolated permit inputs which are maskable. The units were designed to use single mode or multimode fibre for connection to a fibre abort link. The concentrator units are FPGA based and can be programmed to execute any logic needed. The design demonstrated the flexibility to extend the system and integrate with the larger accelerator complex while achieving the response require for PIP-II.

## PLANS FOR PIP-II MPS

MPS operational experience at PIP2IT was useful for evaluating some hardware, firmware algorithms, and aspects of the architecture needed for the PIP-II project. The development and subsequent testing of differential beam current monitoring verified the algorithm and the implementation for this system. The method will be reemployed on newer hardware for transmission losses with AC-CTs at PIP-II. Ring Pickup electrodes and the methods implemented to monitor chopped beam and losses in the MEBT will be designed into the version. The WFE MPS architecture will be for the most part re-implemented on a modular FPGA platform as part of the larger distributed system. The Concentrator hardware with some minor firmware modification to accommodate the specific of its application will be reused in conjunction with other hardware. The Configuration and Mode control experience gained at PIP2IT will be utilized to develop robust monitoring tools. The final overall architecture of the distributed MPS will be designed on the FPGA based the concentrator experience gained as the response times introduced and measure proved adequate for the needs ahead.

## ACKNOWLEDGEMENTS

The authors are thankful to the many people who contributed to building the PIP2IT and those who contributed to its operation and MPS commissioning, these include but are not limited to, D. Crawford, A. Feld, M. Geelhoed, B. Hendrix, K. Koch, A. Prosser, G. Saewert, T. Zuchnik.

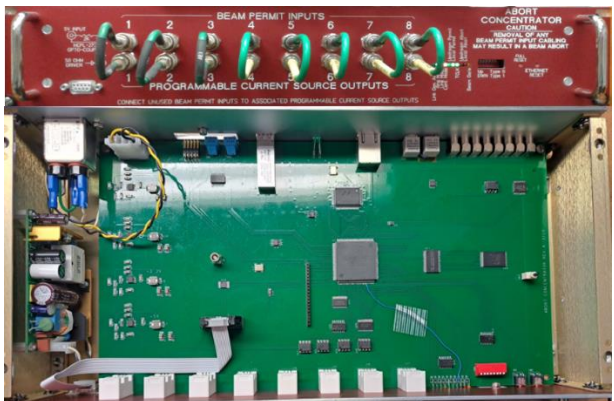

Figure 7: Eight Channel Signal Concentrator box front panel and interior views